\documentclass[10pt,a4paper]{article}
\usepackage{f1000_styles}

\usepackage[numbers]{natbib}
\usepackage{mhchem}
\usepackage{color}
\usepackage[normalem]{ulem}
\usepackage{siunitx}
\usepackage{dcolumn}
\usepackage{comment}
\newcolumntype{d}[1]{D{.}{.}{#1}}

\usepackage{physics}
\usepackage{xparse}

\newcommand{\mbf}[1]{\ensuremath{\mathbf{#1}}}

\NewDocumentCommand{\rep}{s d<| d|>}{%
\IfBooleanTF{#1}{
   \IfValueTF{#2}{
       \IfValueTF{#3}{\braket{#2}{#3}}{\bra{#2}}
       }{
       \IfValueTF{#3}{\ket{#3}}{}
       }
   }{
   \IfValueTF{#2}{
       \IfValueTF{#3}{\braket*{#2}{#3}}{\bra*{#2}}
       }{
       \IfValueTF{#3}{\ket*{#3}}{}
       }
   }
}

\NewDocumentCommand{\rbra}{sm}{\IfBooleanTF{#1}{\rep*<#2|}{\rep<#2|}}
\NewDocumentCommand{\rket}{sm}{\IfBooleanTF{#1}{\rep*|#2>}{\rep|#2>}}
\NewDocumentCommand{\rbraket}{smm}{\IfBooleanTF{#1}{\rep*<#2||#3>}{\rep<#2||#3>}}

\NewDocumentCommand{\field}{o m e{_} e{^} o e{_} e{^}}{
\IfValueTF{#5}{\overline{
  #2\IfValueT{#3}{_#3}\IfValueT{#4}{^{\otimes #4}} %
  \otimes 
  #5\IfValueT{#6}{_#6}\IfValueT{#7}{^{\otimes #7}} %
  \IfValueT{#1}{;#1}
}}{
  \IfValueTF{#4}{\overline{
     #2\IfValueT{#3}{_#3}\IfValueT{#4}{^{\otimes #4}}
     \IfValueT{#1}{;#1}
  }}
  {#2\IfValueT{#3}{_#3}}
}
}

\NewDocumentCommand{\frho}{o e{_} e{^}}{
\field[#1]{\rho}_{#2}^{#3}
}
\NewDocumentCommand{\fdelta}{o e{_} e{^}}{
\field[#1]{\delta}_{#2}^{#3}
}

\newcommand{\e}{a}  %

\newcommand{\br}{\mbf{r}}
\newcommand{\bx}{\mbf{x}}
\newcommand{\bxhat}{\hat{\mbf{x}}}

\NewDocumentCommand{\ex}{e_}{
\IfValueTF{#1}{\e_{#1}\bx_{#1}}{\e\bx}
}  %

\NewDocumentCommand{\lm}{e_}{
\IfValueTF{#1}{l_{#1}m_{#1}}{lm}
}
\NewDocumentCommand{\nlm}{e_}{
\IfValueTF{#1}{n_{#1}\lm_{#1}}{n\lm}
}
\NewDocumentCommand{\enlm}{e_}{
\IfValueTF{#1}{\e_{#1}\nlm_{#1}}{\e\nlm}
}
\NewDocumentCommand{\en}{e_}{
\IfValueTF{#1}{\e_{#1}n_{#1}}{\e n}
}
\NewDocumentCommand{\nlk}{e_}{
\IfValueTF{#1}{n_{#1}l_{#1}k_{#1}}{nlk}
}
\NewDocumentCommand{\enlk}{e_}{
\IfValueTF{#1}{\e_{#1}\nlk_{#1}}{\e\nlk}
}
\NewDocumentCommand{\enl}{e_}{
\IfValueTF{#1}{\en_{#1}l_#1}{\en l}
}
\NewDocumentCommand{\nl}{e_}{
\IfValueTF{#1}{n_{#1}l_#1}{n l}
}

\NewDocumentCommand{\nnl}{s}{
\IfBooleanTF{#1}{n_1 n_2 l}{n_1; n_2; l}
}
\NewDocumentCommand{\ennl}{s}{
\IfBooleanTF{#1}{\en_1 \en_2 l}{\en_1; \en_2; l}
}

\NewDocumentCommand{\gslm}{s}{
\IfBooleanTF{#1}{\sigma\lambda\mu}{\sigma;\lambda\mu}
}

\newcommand{\D}[1]{\operatorname{d}{#1}\,}

\newcommand{\Shat}{\hat{S}}
\newcommand{\Rhat}{\hat{R}}

\newcommand{\nmax}{n_\text{max}}
\newcommand{\lmax}{l_\text{max}}

\usepackage{bm,upgreek}

\newcommand{\bJ}{\mbf{J}}
\newcommand{\nfeat}{n_{\text{feat}}}
\newcommand{\nj}{n_{\text{J}}}
\newcommand{\krn}[0]{\operatorname{k}}

\newcommand{\feat}{\upxi}
\newcommand{\bfeat}[0]{\ensuremath{\bm{\upxi}}}

\begin{document}
\pagestyle{fancy}

\title{
 Local invertibility and sensitivity of atomic structure-feature mappings
}

\author[1]{Sergey N. Pozdnyakov}
\author[2]{Liwei Zhang}
\author[2]{Christoph Ortner}
\author[3]{G\'abor Cs\'anyi}
\author[1,4]{Michele Ceriotti}

\affil[1]{Laboratory of Computational Science and Modelling, Institute of Materials, Ecole Polytechnique F\'ed\'erale de Lausanne, Lausanne 1015, Switzerland}
\affil[2]{Department of Mathematics, University of British Columbia, 1984 Mathematics Road, Vancouver, BC, Canada V6T 1Z2}
\affil[3]{Engineering Laboratory, University of Cambridge, Trumpington Street, Cambridge CB2 1PZ, United Kingdom}
\affil[4]{michele.ceriotti@epfl.ch}

\maketitle
\thispagestyle{fancy}

\begin{abstract}

The increasingly common applications of machine-learning schemes to atomic-scale simulations have triggered efforts to better understand the mathematical properties of the mapping between the Cartesian coordinates of the atoms and the variety of representations that can be used to convert them into a finite set of symmetric  {\em descriptors} or {\em features}.
Here, we analyze the sensitivity of the mapping to atomic displacements, showing that the combination of symmetry and smoothness leads to mappings that have singular points at which the Jacobian has one or more null singular values (besides those corresponding to infinitesimal translations and rotations). This is in fact desirable, because it enforces physical symmetry constraints on the values predicted by regression models constructed using such representations. 
However, besides these symmetry-induced singularities, there are also spurious singular points, that we find to be linked to the \emph{incompleteness} of the mapping, i.e. the fact that, for certain classes of representations, structurally distinct configurations are not guaranteed to be mapped onto different feature vectors. 
Additional singularities can be introduced by a too aggressive truncation of the infinite basis set that is used to discretize the representations. 
\end{abstract}

\section*{\color{OREblue}Keywords}

Atomistic simulations, machine-learning, structural descriptors

\clearpage
\pagestyle{fancy}

\section*{Plain language summary}

The transformation of the atomic coordinates of a molecule or a material to a more symmetric mathematical form is the first step in the application of machine-learning techniques to atomic-scale simulations. 
The properties of such mappings -- e.g. how much the symmetric descriptors change upon deformations of the original structure -- have a knock-on effect on any model built on top of it. 
We study some of the most widely used families of descriptors, revealing how both their fundamental nature and the details of how they are implemented affect the sensitivity of the descriptors and thus the accuracy of the subsequent machine-learning models.

\section*{Introduction}

There has been a tidal wave of interest in the last decade in applying machine learning tools to atomistic modelling problems. See for example the recent thematic issue of Chemical Reviews for a collection of review articles\cite{ceri+21cr}. %
In this first ``heroic phase'' of the development of this field, authors used a wide variety of encodings of atomic structure and regression methods to make models. While it was widely recognised that it is advantageous to encode the physical symmetries of translation, rotation and permutation invariance into  {\em descriptors} of atomic structure, there was little enthusiasm (or opportunity) to rigorously evaluate desirable general properties of different descriptors, as well as compare them with one another, independently of the regression methods and specific applications. 

A number of recent papers have taken on such challenges. It is now understood that many of the descriptors based on the local neighbourhood density are equivalent in the limit of high resolution\cite{musi+21cr} and can be derived from body-ordered expansions of a suitably defined atomic density\cite{will+19jcp,drau19prb}.  

In the present paper we will continue our theoretical investigation of representations of {\em local atomic environments}, $A_i$, given in terms of a feature vector $\bfeat = \bfeat(A_i) = \{\feat_q(A_i)\}_{q=1\ldots\nfeat}$ that is invariant under rotations, reflections and permutations of like atoms. 
Such a representation immediately leads to the question whether the atomic environment $A_i$ can be reconstructed from the features $\bfeat$, up to symmetries. In the context of representing atomic environments (and global structures) this was first explored in some detail in \cite{pozd+20prl} where it was immediately observed that the invariance of the features under said symmetries makes this a formidable theoretical challenge. 
For example, it was shown that any descriptor based on 3-body features cannot uniquely identity \emph{every} configuration containing four or more neighbours, while there are configurations with seven or more neighbors that cannot be distinguished by any descriptor based on 4-body features. That is, pairs of atomic environments can be constructed that are indistinguishable under these descriptors. These observations apply to the vast majority of descriptors used in the field, including in particular  \cite{behl-parr07prl,bart+10prl}; see \cite{musi+21cr} for an extensive discussion.

This challenge points to the fundamental question under which conditions the feature vector $\bfeat$ is a {\em coordinate system}, or in other words, whether its image is a smooth manifold. Since regression, classification and reconstruction tasks are primarily undertaken in feature space, this would be a highly desirable property for the performance of algorithms employed in such tasks. Aside from the injectivity of $\bfeat$ alluded to in the previous paragraph, and which encapsulates the global structure, the immediate next question is to understand its {\em local} structure. 
That is, we will investigate whether $\bfeat$ provides a local coordinate system, i.e. is locally smooth and invertible.
This can be adressed by studying the {\em sensitivity} of $\bfeat$ through properties of its Jacobian matrix. 
In the context of atomic structures, similar studies were first undertaken in Ref.~ \cite{pars+20mlst,onat+20jcp}, where the sensitivity of $\bfeat$ was related  to the accuracy of machine-learning models based on the features, and in Ref.~\cite{parsaeifard2021arxiv}, where it was used to identify regions with near-constant value of the features $\feat_q$.

The purpose of the present work is to explore in more detail the issues of sensitivity, local invertibility, and stability of descriptors constructed from symmetrised $\nu$-correlations of the local atomic density. 
In high symmetry configurations a {\em natural} loss of sensitivity occurs for all smooth and invariant features (we explain this in detail in the main text), which may even be beneficial for regression tasks. A more disturbing observation, analogous to the degenerate pairs discovered in Ref.~\cite{pozd+20prl}, is that certain popular descriptors also exhibit a loss of sensitivity in {\em non-symmetric} configurations. 
We will demonstrate that spurious singularities (loss of sensitivity in non-symmetric structures) arise by two mechanisms: (1) a lack of numerical resolution in the discretisation of the atomic density, which is of course easily remedied; and (2) as intersections of degenerate pair manifolds, which are more fundamental and non-trivial to remove or rule out.
\section{Theory}

\subsection{Symmetry-adapted $\nu$-correlations}
Our main focus is on a class of representations that correspond to $\nu$-point correlations of the atom density, $\rho_i(\br)$, expressed relative to an environment $A_i$, a finite neighbourhood centered on the $i$-th atom.
The density can be written, using the notation introduced in Refs.~\citenum{will+18pccp,will+19jcp}, and formalized in Ref.~\citenum{musi+21cr}, as
\begin{equation}
\rep<\bx||\rho_i> = \sum_{j\in A_i} \rep<\bx||\br_{ji}; g>,
\label{eq:x-rho-i}
\end{equation}
where $\rep<\bx||\br_{ji}; g>\equiv g(\bx-\br_{ji})$ is a Gaussian of width $\sigma_\e$ centered on the vector $\br_{ji}=\br_j-\br_i$ that separates the central atom $i$ and its $j$-th neighbor, with $j$ running over the indices of atoms within $A_i$.
The symmetrized $\nu$-point correlations are obtained by integrating over rotations (and inversion) of tensor products of this density
\begin{equation}
\rep<\bx_1; \ldots \bx_\nu||\frho_i^\nu> = 
\int \D{\Rhat} \rep<\bx_1|\Rhat\rep|\rho_i> \cdots \rep<\bx_\nu|\Rhat\rep|\rho_i>.
\label{eq:x-rho-i-nu}
\end{equation}

These definitions are quite abstract, but encompass a majority of the representations that have been used in the application of machine learning to atomistic problems -- including atom-centered symmetry functions\cite{behl-parr07prl,smit+17cs,zhan+18prl}, smooth overlap of atomic positions (SOAP) powerspectrum\cite{bart+13prb}, bispectrum \cite{bart+10prl}, FCHL descriptors\cite{fabe+18jcp}, all of which are limited to low correlation orders, as well as representations for which $\nu$ can be increased systematically, such as the moment tensor potential (MTP)\cite{shap16mms}, the atomic cluster expansion (ACE)\cite{drau19prb} and the $N$-body iterative contraction of equivariants (NICE)\cite{niga+20jcp}. 

In practical use, the density-correlation is discretised into a feature fector using a basis, and the choice of basis %
can have an impact on how effectively structural information can be stored in the corresponding feature vector (see e.g. Ref.~\cite{gosc+21mlst}). However, we shall first focus on the nature of density-correlation representations in the complete basis set limit. For instance, it was recently shown that three and four-body correlations (corresponding to $\nu=2,3$) are \emph{incomplete}, i.e. it is possible to find pairs of \emph{degenerate} environments, $A_i^+$ and $A_i^-$, that are not related by symmetry, but have the same $\rep|\frho_i^2>$ or  $\rep|\frho_i^3>$ representation\cite{pozd+20prl}. 
It is important to stress that this is true of any descriptor that is a discretized version of these low-order representations, and that non-linear models built on top of such descriptors, no matter how complicated or sophisticated,  cannot eliminate this fundamental shortcoming.

\subsection{Overlap matrix representations}
We also consider a second class of representations of structures that are derived from the eigenvalues of distance matrices~\cite{zhu+16jcp}, and which are far less well understood both theoretically and in applications. 
Because the distance matrix is non-linearly transformed into a form that resembles orbital overlap matrices from quantum chemistry, this representation is usually referred to as ``overlap matrix fingerprints'' (OMFP). 
It was observed numerically that OMFPs are able to distinguish the degenerate atomic environment pairs identified in Ref. \cite{pozd+20prl}, that are indistinguishable by low-order correlation features. To the best of our knowledge, however, no theoretical framework exists to explain this, nor any results suggesting that this is a general property and that no degenerate structures exist for OMFP descriptors. 

To construct a set of OMFP, one begins by specifying an  overlap matrix, ${\bf T}$, analogous to a non-orthogonal tight-binding model with $N_{\rm orb}$ {\it artificial} orbitals per atom. For an atomic environment $A_i$ that has $N$ atoms,  $\bf T$ has $N \times N_{\rm orb}$ rows and columns, and has blocks 
\begin{equation}
	T_{jj'} = f_{\rm cut}(r_{ij}) f_{\rm cut}(r_{ij'}) t(\br_{jj'}) \in \mathbb{C}^{N_{\rm orb} \times N_{\rm orb}},
\end{equation}
where $t$ is some non-linear transformation of the interatomic distances. The OMFP is then the ordered spectrum $\{\tau_k\}_{k = 1}^{N \times N_{\rm orb}}$ of ${\bf T}$. 
If ${\bf T}$ is covariant to rotations then the spectrum will be invariant. In our computational experiments in the present work we will use this form, as proposed in \cite{zhu+16jcp}, in particular using $N_{\rm orb} = 4$ (s and p orbitals). In order to ensure consistency, we compute OMFP with the same implementation  used in previous studies  \cite{pars+20mlst, zhu+16jcp}.

Due to eigenvalue crossings the spectrum is non-smooth: the derivatives of the ordered set of eigenvalues with respect to atomic positions are discontinuous, and so for most tasks (in particular, regression) one should project it onto a basis, e.g. of polynomials, 
\begin{equation}
	\rep<n||A_i; {\bf T}> := \sum_k (\tau_k)^n = {\rm tr}({\bf T}^n).
	\label{omfppoly}
\end{equation}
More importantly for us, this transformation reveals (see Section 5.3 and Eq. (91) in Ref. \cite{musi+21cr} for the details) that the feature $\rep<n||A_i; {\bf T}>$ can be explicitly written as an $n$-correlation, because it is given by the sum of products of elements of $\bf T$ with $n$ factors.
That is, the OMFP contain \emph{some} of the density correlation features up to correlation order $N$. The open question is to understand whether these have genuinely high correlation order or are actually just high-order polynomials of low correlation-order features such as in the example discussed in Sec. VI.C of Ref. \cite{musi+21cr}.
To the best of our knowledge, all applications of OMFP thus far have used the ordered spectrum directly, hence there is no established choice of a basis to obtain smooth features (using the monomials in \eqref{omfppoly} would lead to numerically instabilities). In what follows we will use, as a proof of concept only, a naive basis of sine and cosine functions
\begin{equation}
\begin{split}
\rep<(2q)||A_i; {\bf T}> := &\sum_k \cos(q\tau_k/K)\\
\rep<(2q+1)||A_i; {\bf T}> :=& \sum_k \sin(q\tau_k/K),
\end{split}
\label{eq:OMFP-sine}
\end{equation}
where $K$ is a parameter of the order of the range spanned by the spectrum.
This enables us to benchmark the impact of the lack of smoothness of the ordered spectrum in our tests. A more careful assessment of different bases is left for future investigation.

\subsection{Sensitivity and the Jacobian}
We want to assess whether the mapping from structure to features results, at least locally, in a {\em coordinate system}, i.e., whether it is possible to relate changes in the features to changes in the structure in a one-to-one manner.  
This question is directly connected to the question of the sensitivity of the features to a deformation of the structure, that has been the subject of recent investigation\cite{pars+20mlst,onat+20jcp}. 
The central quantity that contains the answer to these questions is the Jacobian $\bJ$, that for the environment-centered density $\rep|\rho_i>$ reads
\begin{equation} \label{eq:drho-i-jalpha}
J_{j\alpha, \bx}(\rho_i) = \frac{\partial}{\partial r_j^\alpha}\rep<\bx||\rho_i>=\frac{\partial}{\partial r_j^\alpha}\rep<\bx||\br_{ji}; g> = \frac{x^\alpha - r_{ji}^\alpha}{\sigma_\e^2}  \rep<\bx||\br_{ji}; g> \equiv \rep<\bx||\br_{ji}; \partial_\alpha g>,
\end{equation}
where $j$ enumerates the neighbors of the central atom, $\alpha$ indicates one of the $x$, $y$ or $z$ Cartesian coordinates, and the notation $\partial_\alpha g$ indicates the derivative of a 3D Gaussian with respect to the $\alpha$ direction. %
$J_{j\alpha, \bx}$ is an infinite-dimensional operator, a generalization of the Jacobian matrix, that for a finite feature vector $\bfeat$ of length $\nfeat$ contains $3N$ rows, corresponding to the coordinates of the $N$ neighboring atoms, and $\nfeat$  columns, corresponding to the number of indices that enumerate the features for a given discretization of the representation. 
Note that this is the transpose of the most common definition of the Jacobian, which we chose due to the analogy with a design matrix where each column corresponds to a feature.

The singular values, $s_k \geq 0, k = 1, \dots, 3N$, of the Jacobian matrix (or, operator) $\bJ$ can be obtained as the square root of the eigenvalues of $\bJ \bJ^*$, and identify the principal modes of variation of the representation. These singular values indicate how much the features change when the atoms are distorted by an infinitesimal amount according to the displacement patterns associated with the corresponding singular vector (i.e., eigenvector of $\bJ \bJ^*$).

\subsection{Sensitivity, local invertibility and global invertibility}
Consider a smooth descriptor $\bfeat : \mathbb{R}^{3N} \to \mathbb{R}^d$, $d \geq 3N$, with jacobian $\bJ : \mathbb{R}^{3N} \to \mathbb{R}^{3N \times d}$. 
We use $\mbf{u}\equiv \{\br_{ji}\}_{j=1}^N$ to indicate the Cartesian coordinates of the neighbors within $A_i$.
If $\bJ({\mbf{u}}_0)$ has full rank $3N$, then the image of  $\bfeat$ is a smooth manifold locally around $\bfeat(\mbf{u}_0)$ and the mapping $\bfeat$ is locally invertible around $\mbf{u}_0$. In other words, $\bfeat$ is a coordinate system, locally around $\mbf{u}_0$. We can also say in this case that $\bfeat$ is {\em sensitive to small perturbations}. The degree of sensitivity may be measured in terms of the singular values of $\bJ$. 

By contrast, in \cite{pozd+20prl} we explored the seemingly much more stringent condition of {\em global} invertibility of a descriptor, i.e., injectivity of the mapping $\mbf{u} \mapsto \bfeat(\mbf{u})$ on the set of all admissible configurations (atomic environments).

\subsection{Sensitivity of $\rho_i$ and its discretisation}
\label{sec:sensitivity_rhoi}
Although this work is primarily concerned with sensitivity of symmetry-adapted features, it is nevertheless instructive to first consider the sensitivity of $\rho_i$ itself. We will briefly summarize the effect that the smearing $\sigma_\e$ and the discretisation have on the sensitivity of $\rep<\bx||\rho_i>$, which highlight important  considerations for practical implementation. For the sake of clarity of presentation we will consider the specific case when $g$ is a Gaussian, but our analysis applies whenever $g$ is analytic and rapidly decaying, i.e. a ``smeared Dirac delta''.

To that end, we compute the scalar product of two rows in the Jacobian of the neighbour density~\eqref{eq:drho-i-jalpha}:
\begin{equation} \label{eq:J-entry}
\int\D{\bx} \rep<\partial_{\alpha'}g;\br_{j'i}||\bx>\rep<\bx||\br_{ji}; \partial_\alpha g> = 
\frac{e^{-\frac{(\br_{ji}-\br_{j'i})^2}{4\sigma_\e^2}}}{(4\pi \sigma_\e^2)^{3/2}}
\frac{2\sigma_\e^2 \delta_{\alpha\alpha'} - (r_{j'i}^{\alpha'} - r_{ji}^{\alpha'})(r_{j'i}^{\alpha} - r_{ji}^{\alpha}) }{4 \sigma_\e^4}.
\end{equation}
In the limit in which the contributions to the density from individual atoms do not overlap ($\br_{jj'} / \sigma_\e \rightarrow \infty$), the Gaussian term tends to $\delta_{jj'}$, and the scalar product reduces to $\delta_{jj'}\delta_{\alpha\alpha'}/(16 \pi^{3/2} \sigma_a^5)$. Thus, the rows of $\bJ$ are orthogonal in this limit, and all the singular values of $\bJ$ are equal to $1/\sqrt{16 \pi^{3/2} \sigma_\e^5}$. In particular we obtain that the condition number of the Jacobian ${\rm cond}(\bJ(\rho_i)) \to 1$, which is indeed the strongest notion of sensitivity we can hope for. 

In other words, in this regime where all atoms do not overlap relative to the smearing parameter $\sigma_a$, $\rep|\rho_i>$ is equally sensitive to the displacement of each neighbour, independent of the structure considered. 

In practical simulations, the density $\rho_i$ will be discretized by projecting it onto an orthogonal basis $\{ \phi_q \}$,
\[
	\rep<q||\rho_i> = \int \phi_q({\bx}) \rep<\bx||\rho_i> \D{\bx}
				= \sum_j \int \phi_q(\bx) g(\bx - \br_{ij}) \, \D{\bx},
\]
and denote the finite feature vector with $\bfeat = \{\feat_q\}_{q=1}^{\nfeat} =  \{\rep<q||\rho_i>\}_{q=1}^{\nfeat}$ where $q$ ranges over a finite index-set. We must now consider whether the perfect sensitivity of $\rho_i$ (in the regime $\br_{jj'} / \sigma \to \infty$) survives under this discretization. 
A straightforward calculation yields 
\begin{equation}
	J_{q,j\alpha}(\bfeat) 
	= \frac{\partial \rep<q||\rho_i>}{\partial r_j^\alpha} 
	= \int \frac{\partial \rho_i}{\partial r_j^\alpha} \phi_q(\bx) \D\bx
	= \int \frac{\partial g(\bx-\br_{j})}{\partial r_j^\alpha} \phi_q(\bx) \D\bx \equiv \rep<q||\br_{ji}; \partial_\alpha g>  .
\end{equation}
Observe that $J_{q,j\alpha}$ are the $L^2$-projection coefficients of $\frac{\partial \rho_i}{\partial r_j^\alpha}$, hence it is natural to consider the projection, $\Pi$, over the \emph{finite} basis set 
\begin{equation}
	\Pi \bigg[ \frac{\partial \rho_i}{\partial r_j^\alpha} \bigg](\bx)
	:= 
	\sum_{q = 1}^{n_{\rm feat}} \rep<q||\br_{ji}; \partial_\alpha g>\phi_q(\bx)
	=
	\sum_{q = 1}^{n_{\rm feat}} J_{q,j\alpha}(\bfeat) \phi_q(\bx).
\end{equation}
With this definition we obtain that we can rewrite the 
inner product of two rows of $J(\bfeat)$, which is now a sum over the features, as 
\begin{equation}
	(J J^*)_{j\alpha, j'\alpha'} 
	= 
	\sum_{q = 1}^{n_{\rm feat}} J_{q,j\alpha} J_{q,j'\alpha'}^*
	=
	\int \Pi \bigg[ \frac{\partial \rho_i}{\partial r_j^\alpha} \bigg]
		  \Pi \bigg[ \frac{\partial \rho_i}{\partial r_{j'}^{\alpha'}} \bigg],
\end{equation}
which is now identical to \eqref{eq:J-entry} except for the projection error. Standard approximation theory results (e.g., \cite{Powell1981-bg}) imply that this error will decay with a rate that depends on the smoothness of $\frac{\partial \rho_i}{\partial r_j^\alpha}$, i.e., on the smoothness of $g$, and on the choice of basis $\phi_q$. 

For simplicity let us assume that $g$ is analytic (though a Gaussian would be entire and yield even stronger results), and that $\phi_q$ is a basis of polynomials which is the most common choice in this field ($\phi_q(x\bxhat) = R_n(x) Y_l^m(\bxhat)$), then the error will be exponentially small. Taking into account also the rescaling of space via the smearing width $\sigma_\e$, and the domain encoded in the cut-off radius, one can obtain that 
\begin{equation}
	\Big\| \Pi \bigg[\frac{\partial \rho_i}{\partial r_j^\alpha} \bigg] - 
				\frac{\partial \rho_i}{\partial r_j^\alpha}  \Big\| 
	\lesssim 
	\exp\big( - c \sigma_\e \min(\nmax, \lmax) \big),
	\label{eq:j-convergence-nl}
\end{equation}
for a constant $c$ that depends on $g$ and has unit of inverse distance. From this, we can conclude again that ${\rm cond}(\bJ(\bfeat)) \to 1$ at an exponential rate as the discretisation parameter {$\min(\nmax,\lmax)$} increases. Recalling the condition number estimate in terms of smearing width \eqref{eq:J-entry} and a brief argument detailed in the Supporting Information yields the combined estimate  
\begin{equation} \label{eq:combined estimate}
	{\rm cond} (\bJ) - 1 \lesssim {e^{- \frac{r_0^2}{8 \sigma_\e^2}}} 
								+ e^{- c \sigma_\e \min(\nmax, \lmax)},
\end{equation}
where $r_0 = \min_{j,j'} r_{jj'}$. 
 These estimates are illustrated, confirmed, and explored more quantitatively in Figure~\ref{fig:cn-basis}.

While \eqref{eq:J-entry} is exact, experience from approximation theory is that our estimates are close to sharp. This strongly suggests that to obtain a discretisation of $\rho_i$ one must first choose $\sigma_\e$ such that {$r_0 / \sigma_\e \gg 1$}, and then choose the discretisation parameters such that $\min(\nmax, \lmax) \gg 1/(c\sigma_\e)$. In this regime, we expect that ${\rm cond}(\bJ(\bfeat))$ will be close to one. 
However if these requirements are not satisfied then we would expect poor sensitivity encoded in the fact that singular values of $\bf J$ will be close to zero. More generally, \eqref{eq:combined estimate} strongly hints that there is an optimal balance between the smearing and discretisation parameters: given a smearing width $\sigma_\e$ there is a minimal resolution that is required but increasing it may not produce a descriptor with more uniform sensitivity. 
Vice versa, given a minimal distance $r_0$ and a resolution $\min(\nmax, \lmax)$ of the basis, there is an optimal choice of smearing width that minimizes the condition number. See in particular the right-hand panel of Figure~\ref{fig:cn-basis} for a quantitative illustration of this effect. 

For the remainder of the theoretical discussion we shall assume that $\sigma_\e$ is chosen sufficiently small and {$\min(\nmax, \lmax)$} sufficiently large so that the resolution of $\rho_i$ will not affect the results. 

\subsection{Loss of sensitivity after symmetrisation}
The case of the symmetrized density correlations is more complicated. Given that $\rep|\frho_i^\nu>$ features are invariant with respect to rotations, the Jacobian has three singular values associated with a rigid rotation of the environment. 
When investigating the singular behavior of $\bJ$ it is then useful to work in a basis of atomic displacements that removes rigid rotations. The translational symmetry of $\rep|\rho_i>$, which is carried over to $\rep|\frho_i^\nu>$, is taken care of by discarding from the Jacobian the row associated with the central atom $i$.
If one considers $\nu=1$ features that only depend on the distances of the neighbours to the central atom, it is clear that perturbations of the structure in which each neighbor is moved without changing $r_{ji}$ will not result in a change of the feature values, and that $\bJ$ has at most $N$ non-zero singular values, reflecting the fact that a 2-body description of the environment is highly incomplete. 

In the remainder of the paper we perform a numerical and geometric analysis of the behavior of $\bJ$, focusing in particular on $\nu=2$ features, corresponding to distances and angles at the central atom. We will see that in certain symmetric configurations all symmetric features will have ``degenerate'' directions corresponding to singular values $s_k = 0$ -- and that this may in fact be beneficial in terms of using these features to learn structure-property relations. However, more importantly, we will construct examples of configurations without such natural symmetries where 2-correlations still have degenerate directions and thus any descriptor based on 2-correlations inherits this degeneracy. In particular this means that the descriptor does not define a local coordinate system which can have severe consequences for reconstructing atomic environments and for regression tasks.

We say that these structures or environments are {\em linearly degenerate}, or {\em linearly degenerate singularities}, to distinguish them from the discrete pairs of degenerate structures discussed in Ref.~\citenum{pozd+20prl}. By contrast, when a zero singular value is caused purely by a symmetry in the structure we will call such a structure a {\em symmetric singularity}.

\subsection{Pedagogical Example}
As a pedagogical example consider the case of a single particle on the real axis, described by its coordinate $x$. Suppose we are interested in properties of this particle that are invariant under reflection, i.e. $O(1)$ symmetry. In this case we may decide to choose $\feat = x^2$ as a feature (or, 1-dimensional feature vector) describing the position. Clearly, knowledge of $\feat$ allows us to reconstruct $x$ up to a sign (the reflection), even near $x = 0$ where the feature $\feat$ is singular, i.e. $\partial_x \feat = 0$. This is a {\em symmetric singularity} since it is induced by the symmetry group and generic to {\em all} symmetric functions. In that point (and only in that point) we cannot even invert the descriptor locally, i.e., $\feat$ does not supply a local coordinate system. However, if $y$ is a reflection-symmetric property, then its Taylor expansion,
\[
	y(x) \sim c_0 + \frac{c_2}{2} x^2 + \frac{c_4}{4!} x^4 + \dots 
\]
has only even terms, and hence can be expressed as a smooth function of $\feat$, $y(x) = \tilde{y}(\feat)$. This suggests that $\feat$ is well-suited for representing structure-property relationships. 

Now suppose we make the less fortunate choice $\feat = \cos(x)$. By analogy with the power spectrum descriptor, $\feat $ is not injective, i.e. we can find pairs of structures (in fact infinite tuples) that map to the same descriptor value, e.g., if $x_{\pm} := \pi \pm \epsilon$ then $\feat(x_+) = \feat(x_-)$. As this degenerate pair meets at $x_0 := \pi$ we obtain a {\em linearly degenerate singularity} expressed by the fact that $\partial_x \feat(x_0) = 0$. 
Of course we could have immediately seen this root of $\partial_x \feat$. However, we emphasize the intersection of degenerate pairs as it appears to be the {\em generic mechanism} underlying such linear degeneracies even in the much more complex case of symmetry adapted $\nu$-correlations. 

A general symmetric property $y$ of course need not be symmetric about $x_0 = \pi$ and hence has a general Taylor expansion, 
\[
	y(x) \sim a_0 + a_1 (x - x_0) + \frac{a_2}{2} (x - x_0)^2 
		+ \frac{a_3}{6} (x - x_0)^3 + \dots. 
\]	
If $(a_1, a_3, \dots) \neq 0$, it will be impossible to represent $y(x) = \tilde{y}(\feat)$ in a (potentially small) neighbourhood of the degenerate point $x_0$.

\begin{figure}[btp]
\centering
\includegraphics[width=1.0\textwidth]{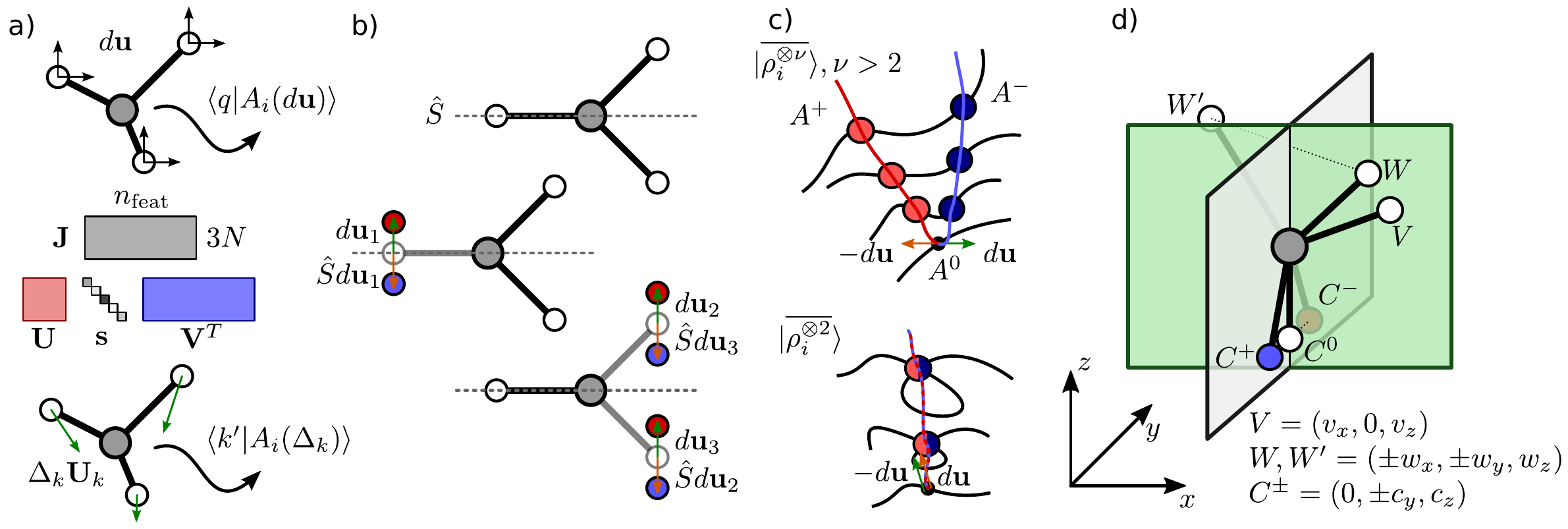}
\caption{\label{fig:symmetry-sketch}
(a) Construction of a basis-set independent sensitivity analysis of the representation: linear perturbations of the feature vector are described by a Jacobian $\bJ$ that couples changes in the $3N$ Cartesian coordinates to changes in the features; right singular vectors project the basis-dependent features into a $3N$-dimensional space that describes the intrinsic sensitivity of the representation.
(b) Examples of a symmetric environment, and two deformations that lead to symmetric pairs of structures, and should therefore be associated with a zero singular value. 
(c) Schematic of the origin of spurious zero singular values associated with pairs of degenerate environments $A^\pm$: if the twin manifolds meet, their intersection produce a ``degenerate singular point'' $A^0$. A small deformation that moves along $A^\pm$ leads to finite separation (and finite feature derivatives) in a non-degenerate feature space, but not when using an incomplete representation.
(d) An example of the construction of a degenerate singularity $A^0$ based on the same family of degenerate structures discussed in Ref.~\citenum{pozd+20prl}. }
\end{figure}

\section{Symmetries and singularities of density-correlation features}\label{sec:singularities}
We now proceed to demonstrate the general concepts we introduced in the previous section for an actual atomistic system, and for a (highly converged) discretization of different classes of representations. 
As an exemplar system we use C-centered environments of  \ce{CH4} configurations, that were used in Ref.~\citenum{pozd+20prl} to construct concrete realizations of the degeneracies that are observed for low-body-order density correlation features. 
We consider different types of configurations to illustrate the various cases in which null singular values of the Jacobian can appear. We compare $\nu=2$ features, $\nu=3$ features, as well as OMFP and their projection on a smooth basis.
We also plot the change in energy of the system as a function of molecular distortions, as an indication of the behavior one should expect for a typical molecular modeling target.

\subsection*{Directionally-resolved sensitivity analysis}

For a feature vector containing a finite number $\nfeat$ of components, describing an environment with $N$ neighbors, the Jacobian is a $3N\times\nfeat$ matrix. Its singular value decomposition, $\bJ = \mbf{U} \operatorname{Diag}(\mbf{s}) \mbf{V}^T$ identifies the principal modes of variation of the representation. 
$\mbf{s}$ is a vector containing $\nj$ principal values (usually $3N$, assuming the typical case in which $\nfeat\gg 3N$) that indicate how much the various features change when the atoms are distorted by an infinitesimal amount according to the displacement patterns associated with each left singular vector contained in the columns of $\mbf{U}$.
The columns in the matrix $\mbf{V}$ describe what feature distortion pattern is associated with each principal component; note that this construction implies that if $\nfeat\gg 3N$ the right singular vectors $\mbf{V}$ span a subspace of dimension smaller than $\nfeat$, and so there are some changes in the feature vectors that cannot be realised by distortions of the structure. 
A crucial observation is that any orthogonal transformation of the features changes the right singular vectors $\mbf{V}$, but not $\mbf{s}$  or $\mbf{U}$. Thus, in the complete basis set limit, one can characterize the sensitivity of representations in a way that is independent of the choice of basis, and so the analysis we carry out in this section is (largely) independent on the details of the discretization.

If a structure $A_i$ is distorted according to a small Cartesian displacement $d\mbf{u}$, the features change according to
\begin{equation}
d\!\rep<q||A_i> = \sum_k (d\mbf{u}\cdot \mbf{U}_k) s_k V_{qk}.
\end{equation}
The magnitude of change of a feature indexed by $q$ is given by the projection of the Cartesian displacement on the left singular vector $\mbf{U}_k$, scaled by the associated singular value $s_k$, and then multiplied by the right singular vectors $\mbf{V}_k$.
Only the latter term depend on the discretization of $\rep|A_i>$, while in the complete basis set limit the left singular vectors $\mbf{U}_k$ and the singular values $s_k$ can be converged with respect to $\nmax$ and $\lmax$. 
Here we use an optimized radial basis\cite{gosc+21jcp} with $\nmax=20$ components, built using the implementation in librascal\cite{musi+21jcp} as the principal components computed based on $1000$ structures from the random \ce{CH4} dataset, and starting from $100$ DVR features, a large angular cutoff $\lmax=20$ and a very sharp density smearing $\sigma_\e=0.05$\,\AA{} to approach the complete basis limit of the density correlation features.
In order to reduce the dependence of the sensitivity analysis on the discretization of the features we do not plot an arbitrary component of $\rep<q||A_i>$, but write the Cartesian displacement around the reference structure in terms of distortions $\Delta_k$ projected along the left singular vectors
\begin{equation}
\Delta\mbf{u} = \sum_k \Delta_k \ \mbf{U}_k(A_i),
\end{equation}
and report on the finite changes of the features projected on the right singular vectors
\begin{equation}
\Delta\rep<k||A_i(\Delta\mbf{u})> \equiv \sum_q V_{qk}(A_i) \left[\rep<q||A_i(\Delta\mbf{u})> - \rep<q||A_i>\right].
\label{eq:rsv-projection}
\end{equation}

Eq.~\eqref{eq:rsv-projection} can also be used when the features that are being probed differ by those that define the left singular directions and $\Delta\mbf{u}$ -- for instance, in what follows we deform the structure along left principal directions for $\rep|A; \frho_i^2>$, but then inspect the change in $\rep|A; \frho_i^3>$ and OMFP. 
It suffices to define the directional derivative of the features along the Cartesian displacement associated with one of the left singular vector, and normalize it
\begin{equation}
\tilde{V}_{qk} = \frac{\partial \rep<q||A_i(\Delta\mbf{u})>}{\partial \Delta_k}; \quad \tilde{\mbf{V}}_k \leftarrow \tilde{\mbf{V}}_k/\left|\tilde{\mbf{V}}_k\right|.
\end{equation}

A subtle but important aspect is that symmetry-invariant features have some ``trivial'' zeros among the singular values, associated with rigid translations and rotations of the environment. 
These singularities should be resolved as a preliminary step, because otherwise non-trivial $s_k=0$ directions would mix with the trivial ones, obfuscating the analysis.
Singularities associated with a rigid translation of an environment are easily eliminated by only considering the Jacobian for the $N$ atoms that are not the environment center. 
Rotations require more attention. A basis of $3N-3$  displacements that are orthogonal to each other and to the displacements corresponding to infinitesimal rigid rotations of the environment around its center can be built based on purely geometric arguments, and the Jacobian should then be projected in this basis. 
The left singular vectors can be converted to build full Cartesian displacements by multiplying them by the transpose of the rotation-less basis matrix. 

\begin{figure}[btp]
\centering
\includegraphics[width=0.9\textwidth]{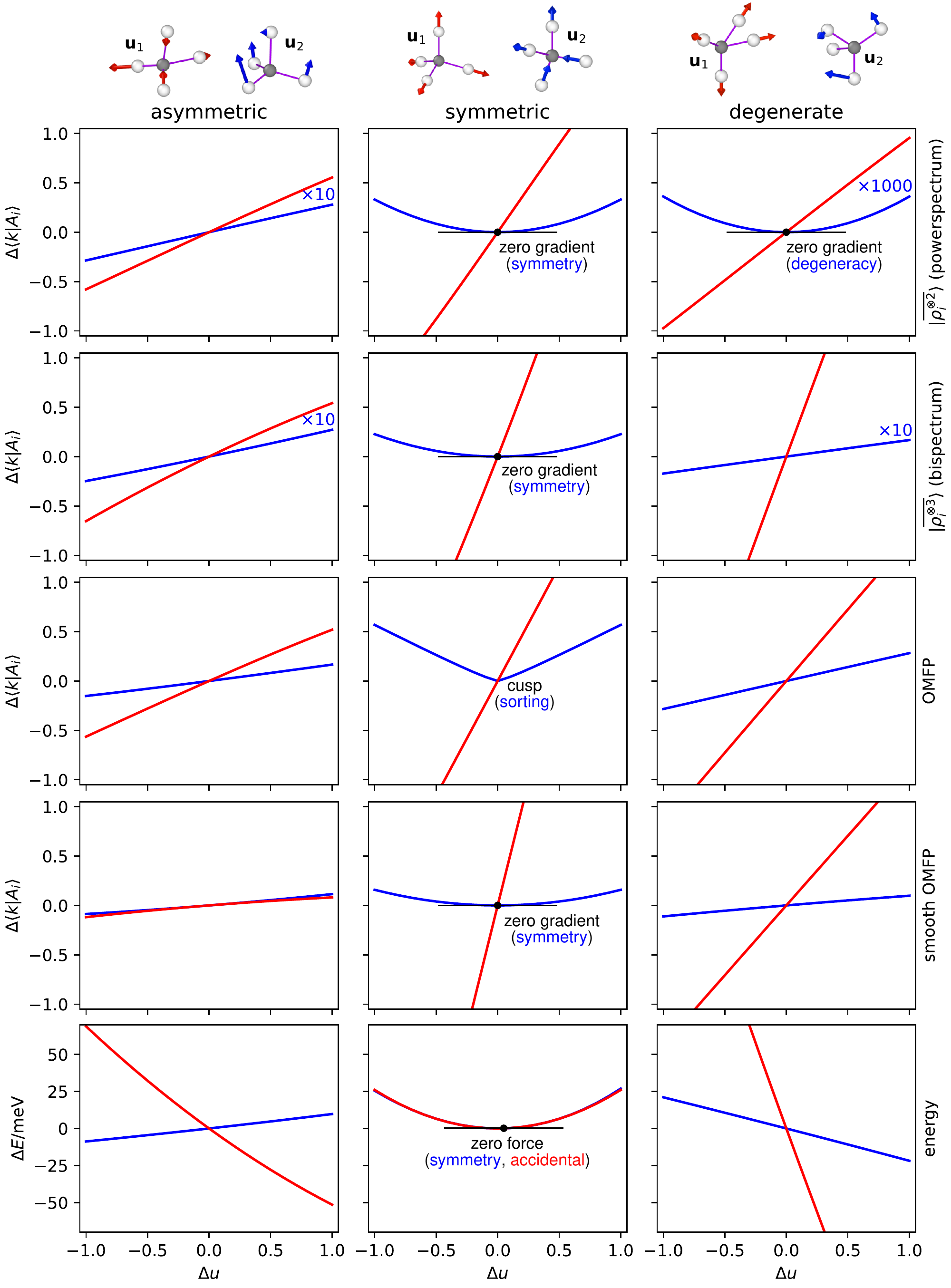}
\caption{\label{fig:fp-response} Changes in the components of features and properties when displacing the H atoms of a \ce{CH4} molecule along selected principal directions.
Results are shown for distortions along two principal directions (determined for $\nu=2$ SOAP features), corresponding to the displacement patterns indicated at the top of the figure, where the color of arrows matches the color of curves.
The columns correspond, left to right, the top and bottom singular values of a distorted, asymmetric \ce{CH4} structure; the symmetric and one asymmetric breathing mode of the ideal methane structure; a ``doubly-degenerate'' structure at the intersection of the two branches of the manifolds discussed in Ref.~\citenum{pozd+20prl}.
Rows show, top to bottom, projected variations of $\nu=2$, $\nu=3$, OMFP and smooth OMFP features, and the total energy of the molecule. %
}
\end{figure}

\subsection*{General structure}
The left column of \autoref{fig:fp-response}  shows the changes in features associated with the largest and smallest singular values of the $\nu=2$ Jacobian for a \ce{CH4} structure corresponding to the optimal tetrahedral geometry, with H atoms distorted at random by $0.3$\AA. 
In the small-displacement regime, the changes in the representation $\rep|A_i>$ are linear, and the slope corresponds to the associated singular value. 
The energy has a non-zero gradient along the two directions, and the relative slope along the two directions reflects, at least qualitatively, the trend seen for all the choices of features except for the smooth OMFP.

\subsection*{Symmetric structure}

Consider an environment $A_i={\br_{ji}}_j$ that is left unchanged by application of a symmetry operation $\Shat$, so that $\Shat A_i = A_i$, with atoms indices mapped as $j\rightarrow \Shat j$, i.e. $\Shat\br_{ji} = \br_{\Shat j i}$ . 
A general infinitesimal distortion of the atoms $d\mbf{u}$ will generate an environment $A_i(d\mbf{u})=\br_{ji}+d\mbf{u}_j$ that is not left invariant by the symmetry. If one can find a displacement pattern such that 
\begin{equation}
\Shat d\mbf{u}_j = -d\mbf{u}_{\Shat j} \label{eq:shat-du}
\end{equation}
then $\Shat A_i(d\mbf{u}) = A_i(-d\mbf{u})$. Any symmetric set of features should be equal for $A_i(d\mbf{u})$ and $A_i(-d\mbf{u})$, meaning that the feature gradient along $d\mbf{u}$ has to be zero: the Jacobian must have one additional singular value equal to zero -- or more if several orthogonal displacements satisfying \eqref{eq:shat-du} can be found. 
A couple of simple examples of these symmetric distortions are shown in \autoref{fig:symmetry-sketch}b.

In the middle column of \autoref{fig:fp-response} we show the sensitivity of different kinds of representations for two types of distortion of a minimum-energy \ce{CH4} structure. The red curve corresponds to a symmetric breathing mode, in which all \ce{CH} bonds are stretched in phase. This corresponds to the only non-zero singular value of the Jacobian for this configuration, and all descriptors change linearly along this direction. 
The blue curve, instead, corresponds to a stretch mode in which two \ce{CH} bonds contract, and two dilate by the same amount. This deformation is symmetric in the sense of Eq.~\eqref{eq:shat-du}, which leads to a symmetric behavior of the descriptors response. 
The case of OMFP merits further discussion: due to the eigenvalue sorting that guarantees atom index permutation invariance, the response has a cusp for $\Delta_2=0$. 
A similar cusp will be present for every symmetric deformation, which in the case of the minimum-energy \ce{CH4}  structure implies all but one degree of freedom. 
Fortunately, it is simple to solve this problem: when using \emph{smooth} OMFP, computed projecting the spectrum on a smooth basis following Eq.~\eqref{eq:OMFP-sine}, the cusp is removed, and the descriptors have zero gradient as required by the combination of symmetry and smoothness. 

A quadratic behavior along the symmetric direction is consistent with the changes in energy, that follow a parabolic trend. For this minimum-energy configuration, the trend is parabolic along \emph{both} directions. For the red curve, however, this quadratic behavior is a consequence of the \ce{CH} bond length being the equilibrium one: if one changed it, or recomputed this same configuration at a different level of theory, there would be a non-zero slope for the energy around $\Delta_1=0$. The blue curve, instead, is a consequence of symmetry, and the trend would be quadratic for this stretch deformation even if the \ce{CH} bonds were not the minimum-energy ones -- provided that they were all equal.

\subsection*{Degenerate structure} 

For $\nu=2$ features (3-body ACSF, SOAP powerspectrum \ldots) one can also find configurations with zero singular values that are not associated with a symmetry, but rather with degenerate structures of the kinds discussed in Ref.~\cite{pozd+20prl}. 
In short, one can find pairs of environments, $A_i^+$ and $A_i^-$, that have exactly the same $\rep|\frho_i^2>$ features, although they are not equivalent and in general have different physical properties. 
If the environments $A_i^\pm$ belong to manifolds $\mathbb{A}_i^\pm$ that cross then this gives rise to a ``doubly-degenerate'' environments $A_i^0$, for which there is at least one direction such that $A_i^0(d\mbf{u}) \in \mathbb{A}_i^+$ and $A_i^0(-d\mbf{u}) \in\mathbb{A}_i^-$. 
Thus, the value of the  $\rep|A^0(d\mbf{u});\frho_i^2>$ features is identical to those of $\rep|A^0(-d\mbf{u});\frho_i^2>$.
As a consequence, the Jacobian $\bJ(A_i^0)$ has a \emph{spurious} zero among its singular values, leading to  quadratic (or higher) variation of $\rep<k||A_i(\Delta\mbf{u})>$ when displacing the atoms along the coordinate associated with it. 
This behavior is very clear in the right column of \autoref{fig:fp-response}.
For a structure sitting at the intersection of the $\mathbb{A}_i^+$ and $\mathbb{A}_i^-$ manifolds, $\nu=2$ density features have a spurious zero singular value, and the corresponding component of $\rep|\frho_i^2>$ change quadratically when displacing the atoms along the left singular vector. 
The pathological nature of this behavior is apparent in the fact that both OMFP and the $\nu=3$ density representation vary with non-zero slope along the same geometric deformation, and so does a physical property such as the molecular energy. 
This shortcoming of low-body-order features is a direct consequence of the lack of injectivity discussed in Ref.~\cite{pozd+20prl}. 

Similar to the order-zero degeneracies discussed in Ref.~\cite{pozd+20prl}, information on the features centered on other atoms may be sufficient to break the tie, and lift the singularity, resulting in non-zero gradient of models of molecular properties. 
Any set of discrete degenerate structures can generate a spurious singularity whenever the degenerate manifolds cross. This is also true for the construction, discussed in Ref.~\cite{pozd+20prl}, of a pair of structures that have the same $\nu=3$ features -- and so also the bispectrum, and similar four-body features, can have linearly degenerate structures.
We could not determine -- but cannot rule out -- the existence of additional manifolds, or isolated structures, which have a spurious zero in the eigenspectrum of $\bJ$ and are not the result of the intersection of those associated with discrete degeneracies.

\begin{figure}[bt]
\centering
\includegraphics[width=0.8\textwidth]{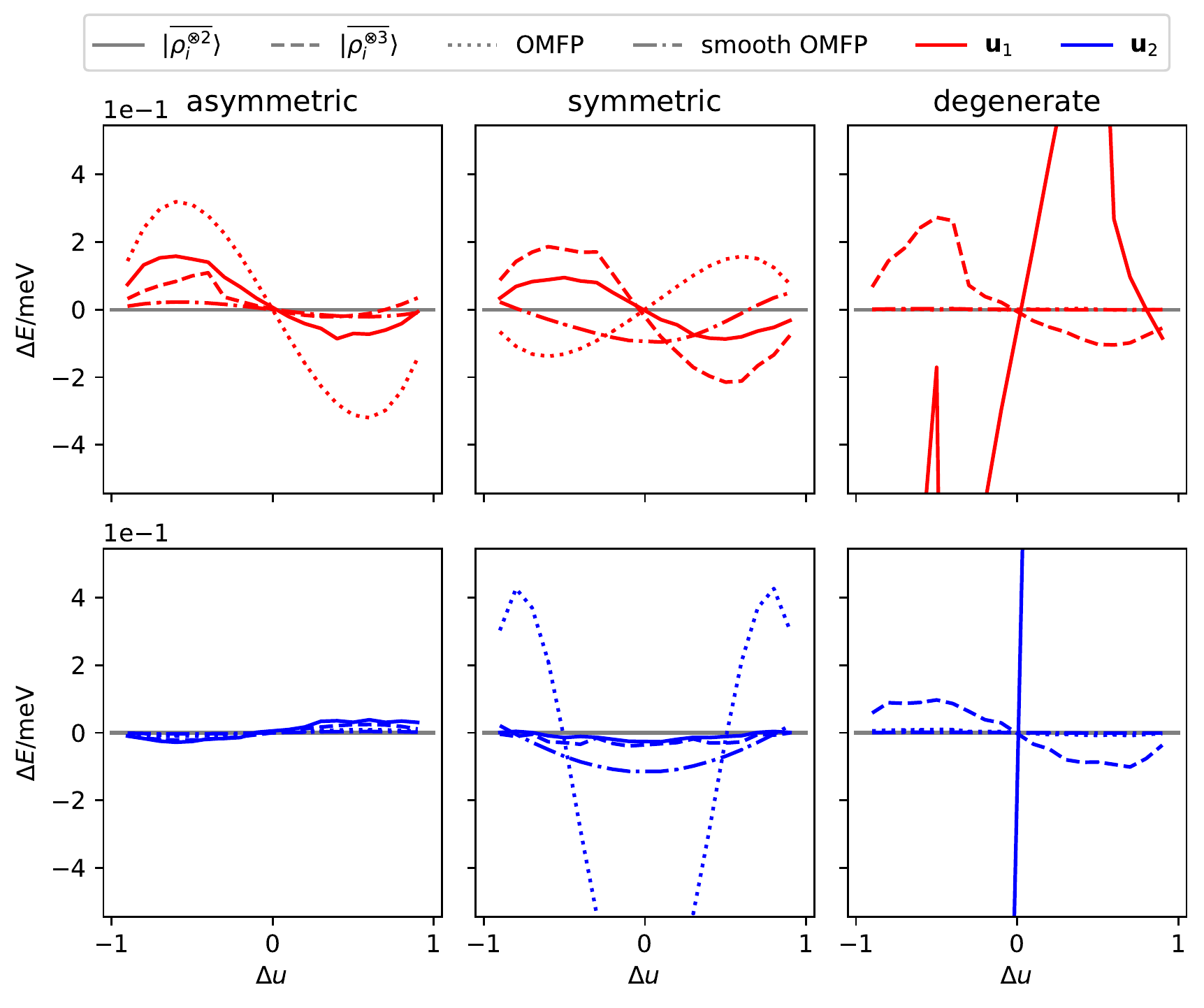}
\caption{\label{fig:fp-errors} Error in fitting the energy of a \ce{CH4} molecule when the atoms are displaced along the directions show in \autoref{fig:fp-response}. All models are trained using a $3\times 3$ grid of points along $\mbf{u}_1$ and $\mbf{u}_2$, and the line dashing indicates the features used for the model. The two rows correspond to cuts along the directions corresponding to the two separate components. 
}
\end{figure}

\begin{table}[b]
\hrule height 0.05cm  \vspace{0.1cm}
\caption{\label{tab:models-accuracy}Accuracy of regression models for the energy across a grid of $21\times 21$ points along the two selected directions, as a function of the regular $n\times n$ sub-grid used for training.
The same structures, distortions and representations are used as those shown in \autoref{fig:fp-response}. Errors are given in $\mu$eV (such small values are possible because we are effectively fitting a 2D function with a range of a few 10s of meV) and correspond to root mean square errors on the test points. 
}
\centering
\begin{tabledata}{ld{3.2}d{3.2}d{3.2}d{3.2}}
\header Structure and grid & \multicolumn{1}{c}{$\rep|\frho_i^2>$} & \multicolumn{1}{c}{$\rep|\frho_i^3>$} & \multicolumn{1}{c}{OMFP} & \multicolumn{1}{c}{s-OMFP}  \\ 
\row asymmetric, $3\times 3$ & 89 &  48 & 230  & 16  \\ 
\row asymmetric, $7\times 7$ & 0.3 & 5.6 & 0.06 & 0.1 \\ 
\hline
\row symmetric, $3\times 3$ & 49 &  140 & 570  & 48  \\ 
\row symmetric, $7\times 7$ & 5.8 & 5.4 & 37 & 5.3\\ 
\hline
\row degenerate, $3\times 3$ & 9600 & 150  & 6.5  &  1.4 \\ 
\row degenerate, $7\times 7$ & 5700 & 0.42 & 0.11 & 0.06\\ 
			
\end{tabledata}
\end{table}

\section{Accuracy of regression models}

To investigate the relationship between the mathematical properties of the structure-feature mapping and the behaviour of a regression model that use them, we fit the energy of the example \ce{CH4} configurations shown in \autoref{fig:fp-response}. 
We chose Kernel Ridge Regression (KRR) based on a squared exponential kernel as a simple model with universal approximation properties. 
The elements of the kernel matrix between training configurations are defined as:
\begin{equation}
    K_{ii'} = \krn(A_i, A_i') \equiv e^{-\gamma \sum\limits_q (\feat_q(A_i)- \feat_q(A_i'))^2}\text{,}
\end{equation}
where $\feat_q(A_i)$ indicates one of the entries in the feature vector associated with an environment $A_i$. 
The representation may be either a discretization of $\ket{\rho^{\otimes 2}}$, $\ket{\rho^{\otimes 3}}$ or a set of OMFP.

The predictions for a new environment $A_\star$ can be written   as 
\begin{equation}
E(A_\star) = \sum_i \krn(A_\star,A_i) b_i, \quad \mbf{b} =  (\mbf{K} + \sigma^2 \mbf{I})^{-1} \mbf{E}
\end{equation}
where $\mbf{K}$ is the kernel between train configurations, and $\mbf{E}$ is the vector of the energies associated with the training structures.
The model is entirely determined by the training set, the features, and two hyperparameters -- the scale of the squared exponential kernel $\gamma$ and the regularization $\sigma^2$. 
We optimize $\gamma$ and $\sigma^2$ by grid search, minimizing the error of the predictions on the structure that are not used for training (further details are given in the Supporting Information). 
Even though this procedure implies some information leaking from the test structures, doing this consistently for all models ensures a fair, and deterministic, comparison. 

The overall errors of the different models are shown in \autoref{tab:models-accuracy}, and the errors along the two one-dimensional cuts associated with the displacement patterns in \autoref{fig:fp-response} are plotted in \autoref{fig:fp-errors}. 
For the generic, asymmetric structure the three representations perform similarly. 
Errors are larger in the direction of $\mbf{u}_1$ -- that also corresponds to the highest singular value for $\rep|\frho_i^2>$ -- than in the direction of  $\mbf{u}_2$ -- that we selected as the lowest singular value. 
For the symmetric \ce{CH4} structure, the global minimum of the methane molecule, the energy model displays a very interesting behaviour. The two deformations correspond to similar displacement-energy curves (\autoref{fig:fp-response}, bottom panel), but the model errors are very different. 
This is because the symmetric breathing mode $\mbf{u}_1$ is an ``accidental'' energy minimum, and so the model predictions yield a small, but non-zero, force for $\Delta_1=0$. 
On the other hand, $\mbf{u}_2=0$ is required to be an extremal point due to symmetry, and so all symmetric features predict, by construction, a symmetric curve with no force for $\Delta_2=0$ and achieve a much lower prediction error. 
The error of the OMFP-based model is considerably higher than the others. This is because of the non-smooth behavior in $\Delta_2 =0$, which means that even if the curve is symmetric, the force in $\Delta_2 \rightarrow 0^\pm $ is not zero.
Indeed, the smooth version of the OMFP has performance comparable to those of the power spectrum.
Finally, displacements around the structure associated with a degenerate singularity result in a potential energy surface that cannot be fitted using $\rep|\frho_i^2>$ features. 
Even though being incapable of reproducing the non-zero force for $\Delta_2=0$ is a serious shortcoming of $\rep|\frho_i^2>$, the effect is dwarfed by the impact of the discrete degeneracies: the model displays an enormous error also for $\Delta_2=\pm 1$,  \emph{even though those displacements corresponds to two of the training points}.
The OMFP model performs better than $\rep|\frho_i^3>$ features, which are however perfectly capable of resolving the degeneracy. 
When increasing the training grid to $7\times 7$ points, all features except for  $\rep|\frho_i^2>$ converge, with only a small residual error.

\section{Finite basis set}\label{sec:basis}

In the discussion this far we have been careful to use a highly-converged discretization of the density-correlation features, so that our numerical results are representative of the intrinsic nature of the representation, rather than of specific parameters, or implementation choices. 
Considerations on the ``physical'' and ``degenerate'' singularities, and the smoothness of the structure-feature mapping, applies equally well to SOAP, ACSF, FCHL,  etc. 
However, implementations details -- in particular the type and size of the basis used to discretize density correlations -- do matter, as they affect the condition number of the Jacobian, possibly introducing numerical instabilities and near singularities. 
This is easy to see by considering an overly parsimonious discretization, in which the number of features is smaller than $3N-6$: then, $\bJ$ has a rank that is insufficient to fully characterize structural distortions. 

\begin{figure}[btp]
    \centering
    \includegraphics[width=1.0\linewidth]{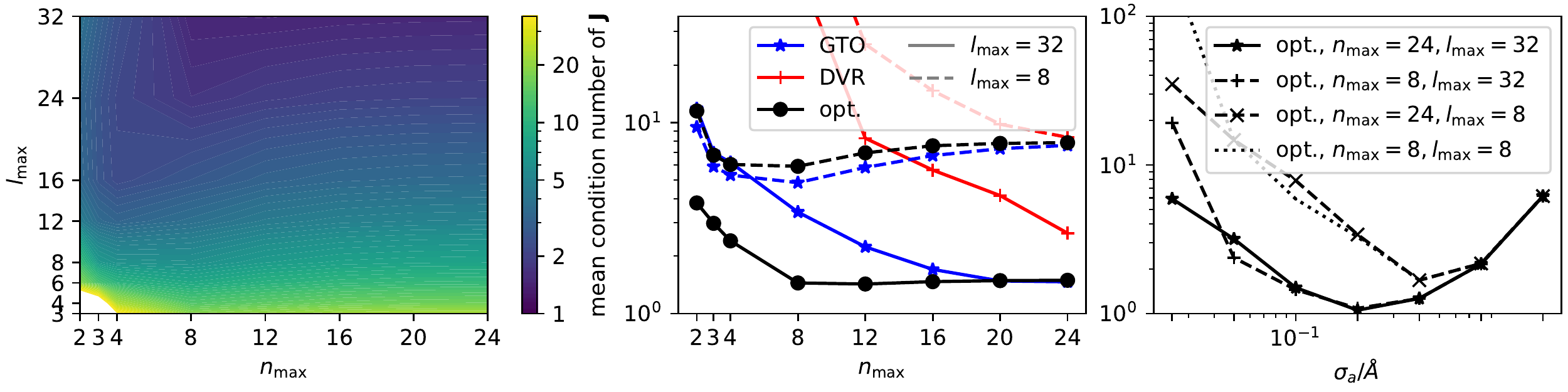}
    \caption{Mean condition number for the Jacobian of $\rep|\frho_i>$ features, computed from random \ce{CH4} configurations\cite{matcloud20a}, for C-centered environments. Left: mean CN as a function of $\nmax$ and $\lmax$, for an optimal radial basis\cite{gosc+21jcp} and a density smearing $\sigma_\e=0.1$\AA. Center: mean CN as a function of $\nmax$, for two selected values of $\lmax$ and comparing three different choices of basis, for $\sigma_\e=0.1$\AA . Right: mean CN as a function of $\sigma_\e$, using an optimal radial basis for different values of $(\nmax,\lmax)$. }
    \label{fig:cn-basis}
\end{figure}

We can then consider a more realistic example, taking a database of \ce{CH4} environments with the H atoms randomly distributed in a sphere of 3\AA{} radius around the central carbon, and assessing the condition number of the Jacobian for the expansion coefficients of the C-centered density, as a function of the basis type and size.
As shown in \autoref{fig:cn-basis}, a large basis set size is needed to approach the theoretical condition number, which should be one for the non-symmetrized density expansion coefficients as discussed in \autoref{sec:sensitivity_rhoi}. 
Furthermore, the nature of the radial basis is very important. A basis optimized to maximize the information content for a given $\nmax$ approaches the limiting value for $\nmax\approx 8$, even though a very large $\lmax$ has to be used, in addition. 
A GTO basis converges more slowly, and the DVR basis leads, for $\nmax=4$, to a mean CN above $10^6$.
These numerical results closely match but quantify more precisely the theoretical predictions made in \S~\ref{sec:sensitivity_rhoi}, and further strengthen the importance of implementation details: even in a case in which one would expect a perfectly-conditioned Jacobian, near-singularities can arise when using a small, or sub-optimal, discretization of the representation.
The right panel of \autoref{fig:cn-basis} demonstrates the role of the Gaussian width as well as its interplay with the discretisation in determining the condition number of $\bJ$. As predicted theoretically by \autoref{eq:combined estimate} the resolution of the individual particles and thus the condition number initially improves with decreasing smearing width, but eventually the numerical discretisation is no longer sufficient to resolve the narrow Gaussian at which point the condition number increases again. The optimal value of $\sigma_\e$ is shifted to the left as the discretisation is refined. 
This analysis supports -- at least in terms of achieving a uniform sensitivity of the density expansion coefficients to atomic displacements -- the choice of a Gaussian smearing of the density that is about half of the minimal interatomic separation, a practice that is often adopted by practitioners when constructing Gaussian approximation potentials\cite{deri+21cr}. It moreover highlights the importance of the discretisation parameters when a small smearing width is employed.

\begin{figure}[btp]
    \centering
    \includegraphics[width=1.0\linewidth]{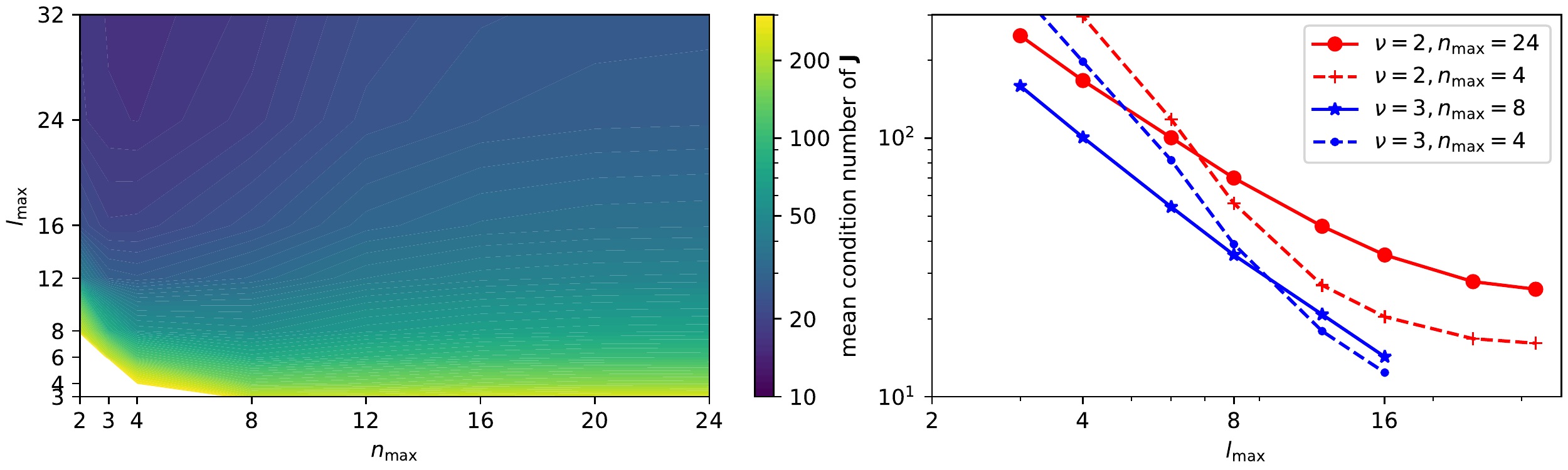}
    \caption{Mean condition number for the Jacobian of $\rep|\frho_i^{\nu}>$ features, computed from random \ce{CH4} configurations\cite{matcloud20a}, for C-centered environments. Left: mean CN as a function of $\nmax$ and $\lmax$, for SOAP ($\nu=2$) features, an optimal radial basis\cite{gosc+21jcp} and and a density smearing $\sigma_\e=0.1$\AA. Right: mean CN as a function of $\lmax$, for different values of $\nmax$, comparing  $\nu=2$ and $\nu=3$ features. }
    \label{fig:cn-soap}
\end{figure}

\begin{figure}[btp]
    \centering
    \includegraphics[width=1.0\linewidth]{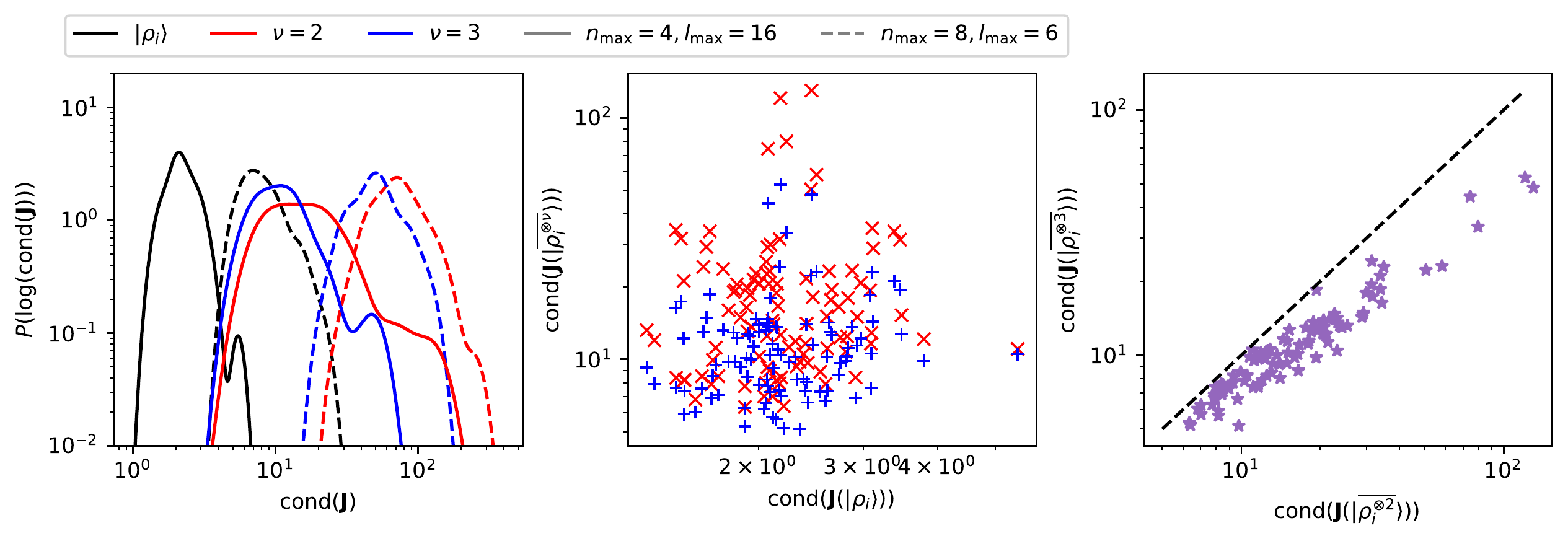}
    \caption{A comparison of the condition numbers of the Jacobian of $\rep|\rho_i>$, $\rep|\frho_i^2>$, $\rep|\frho_i^3>$, computed from random \ce{CH4} configurations\cite{matcloud20a}, for C-centered environments. 
    Left: histogram of the CN of different structures for typical discretization parameters ($\nmax=8, \lmax=6$; dashed lines), and parameters that lead to the best CN for $\nu=3$ features ($\nmax=4, \lmax16$; full lines).  
    Center: parity plot of the condition number for 100 random structures, comparing the values for density expansion coefficients (x-axis), and those for $\nu=2,3$ density correlations (y-axis), using $\nmax=4, \lmax=16$ consistently. 
    Right: parity plot of the condition number for the same 100 random structures, comparing the values for $\nu=2$ (x-axis), and those for $\nu=3$ (y-axis) density correlations. }
    \label{fig:cn-histo}
\end{figure}

Results for density-correlation features are qualitatively similar (which is unsurprising, given that they are computed from combinations of density expansion coefficients) but deserve further comments. As shown in \autoref{fig:cn-soap}, the mean condition number of $\nu=2$ features is higher than of  $\rep|\frho_i>$, and we could not reduce it below 10-20.
At the same time, the  dependence of the condition number on $\nmax$ is sharper than for the density expansion coefficients: the presence of products of radial functions that is implicit in the definition of $\rep<n_1; n_2; l||\frho_i^2>$ implies that fewer starting basis functions are needed to reach a satisfactory description of radial degrees of freedom. 
Similar observations can be made for the $\nu=3$ correlations: the condition number decreases quickly with both $\nmax$ and $\lmax$ -- the correlations involve products of multiple radial and angular terms -- but it is very difficult to reduce the mean condition number below $\approx 10$. 

The question is whether the higher condition number of $\nu=2,3$ invariant features is due to the amplification of the singular values associated with the fact that density correlations are products of the expansion coefficients, or is associated with the proximity to the geometric singularities discussed in \autoref{sec:singularities}.
To elucidate this point, we show in \autoref{fig:cn-histo} an analysis of the condition numbers computed for 100 representative random \ce{CH4} structures. 
For a given discretization, the CN of the symmetrized features is consistently higher than that of the underlying $\rep|\rho_i>$ expansion coefficients. 
However, $\nu=3$ features have consistently a \emph{lower} CN than the powerspectrum features, which indicates that the relationship between body order and the condition number is not as simple as if it reflected just the products that enter the definition of high-$\nu$ representations. 
In addition, $\nu=2,3$ features have a more pronounced high-CN tail. The corresponding structures have a low condition number for $\bJ(\rep|\rho_i>)$ (\autoref{fig:cn-histo}, center), suggesting that the large CN is not a consequence of the discretisation, but is linked to proximity to symmetric (or degenerate) singular points. 
To determine the nature of the singularities we use a similar strategy to that adopted in Ref.~\citenum{pozd+20prl} to study discrete power spectrum degeneracies. 
We compare the CN of $\bJ(\rep|\frho_i^2>)$ against that of $\bJ(\rep|\frho_i^3>)$. 
Structures that approach a degenerate singularity should have a large $\operatorname{cond}(\bJ(\rep|\frho_i^2>))$, and a much smaller $\operatorname{cond}(\bJ(\rep|\frho_i^3>))$. 
As shown in \autoref{fig:cn-histo} (right), all the structure we considered with a high CN for powerspectrum features have also a comparatively high CN for the bispectrum features, which indicates that they are (close to) symmetric singularities, and that none of the 100 random structures we considered are close to a degenerate singularity. 

Overall, this numerical analysis supports the analytical estimates in \autoref{sec:singularities}: the basis set and the density smearing must be converged (and balanced) to obtain neighbor density coefficients with a uniform sensitivity to deformations. 
For some configurations, symmetric representations  exhibit (near) singular behavior, which is, at least for this dataset, consistent with physical constraints and not a manifestation of pathological behavior.

\section{Conclusions}

We have investigated, both analytically and numerically, the sensitivity to structural distortions of a family of symmetric representations of atomic environments that can be interpreted as a hierarchy of $\nu$-point correlations of the neighbour density, and compared it with that of OMFP, an alternative set of features that incorporates \emph{some} of the components at each order of correlation. 
Some of the results we show are intrinsic properties of the mathematical structure that underlies density-correlation representations, and independent of the discrete basis used to compute them as a finite-dimensional vector.
In the limit of a complete basis, and for a sharp density, the mapping from coordinates to \emph{translationally} invariant environment features has uniform sensitivity with respect to the displacement of any of the neighbors, as measured in terms of the Jacobian.
Enforcing $O(3)$ group symmetries, however, introduces some symmetric singular points -- configurations for which symmetry implies that the Jacobian has fewer than $3N-3$ non-zero singular values. 
These symmetric singularities are physically meaningful, and beneficial to the regression of properties of the structures that are bound by symmetry to have critical points for those configurations.  
For low orders of correlation, we also find that the symmetric density correlation features present \emph{degenerate singularities}, that are directly linked to the discrete degenerate structrues that have been recently shown to be a manifestation of the incompleteness of the structure-feature mapping\cite{pozd+20prl}. 
These singularities are unphysical, and detrimental to the construction of energy models -- although potentially less so than the discrete degenerate manifolds they derive from.

It is important to stress that, even if there are many degenerate singularities arranged along a continuous manifold, in all cases we have observed so far the features are \emph{not} constant along that manifold as suggested in~\cite{parsaeifard2021arxiv}. Instead, the direction of the said manifolds only indicates the absence of invertibility of the linearised structure-feature mapping, along one or more orthogonal directions. Higher-order correlations, either from a systematic expansion or from OMFP, cure these problems. In the latter case the lack of smoothness in the feature mapping has a negative impact on the accuracy of models close to symmetric structures, which is easily cured by projecting the spectrum on a smooth basis. 

We also consider how these general results change when using a \emph{finite} discretization of the features. 
We find that the use of a small basis means that even the Jacobian of the mapping between atomic coordinates and neighbor density coefficients -- that ought to have uniform sensitivity -- can have a large condition number, that depends on the type and number of basis functions as well as on the smearing of the atom density, with optimal results obtained for an intermediate smearing of a fraction of the minimum interatomic separation. 
Symmetrized features computed with a limited basis, too sharp or too broad Gaussian smearing inherit the anisotropic response to atomic deformations from the expansion coefficients they are built from. Even when using converged density coefficients, however, configurations with a high condition number can be found, corresponding to structures that are close to one of the symmetric singularities.
We find instead that -- at least in the benchmark dataset we consider -- degenerate singularities are much rarer.

This \emph{local} analysis of the nature of the structure-features mapping complements the geometric arguments of our previous work, and clarifies that the presence of structures or manifolds in which the Jacobian is low rank (or with a very large condition number) can be physical, but also the result of artifacts introduced by the incompleteness of the low-order descriptors or the excessive truncation of the basis used to discretize the density-correlation features. 
While we focus on applications to chemical and materials sciences, our results have obvious implications for any framework based on a description of point clouds, and the increasingly popular equivariant neural network frameworks. 
The challenges we discuss in the present work also appear in other contexts, for example in signal processing in relation to the ``phase recovery'' problem\cite{Yellott1992-gh,kaka12jmiv} (see also \cite{Uhrin2021-ac} for an extensive review) or in distance geometry where a closely related challenge is the ``unassigned distance geometry problem'' \cite{bout-kemp04aam,Duxbury2016-iv}.
From a mathematical perspective, there are several open questions, that are closely related to those that arise for discrete degeneracies. The existence of classes of degenerate singularities different from those we present, the frequency of problematic structures in realistic scenarios, whether and how many degeneracies remain as the body order of the descriptors is increased, and more generally the role of symmetries in changing the topology of the structure-feature mapping are all aspects for which further investigation is needed, and for which this work provides useful conceptual and numerical tools. 

\section*{Data and software availability} %
Data and software to replicate this study is available on public repositories. 

\subsection*{Source data}

The randomly-distorted \ce{CH4} structures used in \autoref{sec:basis} are a random representative subset of those distributed in the following data record: 
If the data has been published previously, details of the dataset and where it can be accessed should be provided here.

\begin{quote}
Repository: Randomly-displaced methane configurations.\\
https://doi.org/10.24435/materialscloud:qy-dp.
\\
\\
This project contains the following underlying data:
\begin{itemize}
	\item methane.extxyz.gz - 7732488 random methane molecules along with their dft energies and forces in the extended xyz format
\end{itemize}

Data are available under the terms of the Creative Commons  Attribution Non Commercial 4.0 International.
\end{quote}

\subsection*{Underlying data}

The data used to generate \autoref{fig:fp-response}, including molecular geometries, reference energies and projected feature displacements, are stored in the following data record. 

\begin{quote}
Repository: Sensitivity benchmarks of structural representations for atomic-scale machine learning.\\
https://doi.org/10.24435/materialscloud:7z-g6
\\
\\
This project contains the following underlying data:
\begin{itemize}
	\item *.chemiscope.json.gz. (2D manifolds corresponding to the plots in \autoref{fig:fp-response}) 
\end{itemize}

Data are available under the terms of the Creative Commons  Attribution Non Commercial 4.0 International.
\end{quote}

\subsection*{Software availability}

Powerspectrum and bispectrum features used in this work were generated using librascal, using the version with commit ID db2e2445d34c196c94731249061740123f9fbc28. librascal can be downloaded freely from  \url{https://github.com/cosmo-epfl/librascal}, under a GNU LGPL 3.0 license.

\section*{Grant information}
MC and SNP acknowledge support from the Swiss National Science Foundation [Project No. 200021-182057] and from the NCCR MARVEL, funded by the Swiss National Science Foundation (SNSF).
This manuscript builds on results that were obtained in the context of the ERC Starting Grant project HBMAP [GA No. 677013]. 
CO is supported by Leverhulme Research Project Grant RPG-2017-191 and by the Natural Sciences and Engineering Research Council of Canada (NSERC) [funding reference number IDGR019381].

\section*{Acknowledgements}
We would like to thank Stefan Goedecker and Behnam Parsaeifard for sharing the code we used to compute OMFPs.

\bibliographystyle{unsrtnat}

\section*{Submitting your article}
Generate a PDF file of your project and submit this alongside a zip file containing all project files (including the source files, style files, and PDF) using our \href{https://open-research-europe.ec.europa.eu/for-authors/publish-your-research}{online submission form}.

\end{document}